\documentclass[prb,twocolumn,showpacs]{revtex4}
\usepackage[paperwidth=210mm,paperheight=297mm,centering,hmargin=2.2cm,vmargin=2.2cm]{geometry}
\usepackage{graphicx}
\usepackage{amssymb,amsfonts,amsmath}
\usepackage{enumerate} 
\usepackage{paralist} 

\begin{document}

\title{Scale invariance of a diodelike tunnel junction}
\author{H.~Cabrera}
\author{D.A.~Zanin}\email{dzanin@phys.ethz.ch}
\author{L.G.~De~Pietro}
\author{Th.~Michaels}
\author{P.~Thalmann}
\author{U.~Ramsperger}
\author{A.~Vindigni}
\author{D.~Pescia}
\affiliation{Laboratory for Solid State Physics, ETH Zurich, 8093 Zurich, Switzerland}
\author{A.~Kyritsakis}
\author{J.P.~Xanthakis}
\affiliation{Electrical and Computer Engineering Department, National Technical University of Athens, Zografou Campus, Athens 15700, Greece}
\author{Fuxiang~Li}
\author{Ar.~Abanov}
\affiliation{Department of Physics, MS 4242, Texas A\&M University, College Station, Texas 77843-4242, USA}
\pacs{73.40.Gk,73.63.-b,68.37.Ef,79.70.+q}
\date{\today}
\begin{abstract}
We measure the current vs voltage ($I$-$V$) characteristics of a diodelike tunnel junction consisting of a sharp metallic tip placed at a variable distance $d$ from a planar collector and emitting electrons via electric-field assisted emission. All curves collapse onto one single graph when $I$ is plotted as a function of the single scaling variable $Vd^{-\lambda}$, $d$ being varied from a few mm to a few nm, i.e., by about six orders of magnitude. We provide an argument that finds the exponent $\lambda$ within the singular behavior inherent to the electrostatics of a sharp tip. A simulation of the tunneling barrier for a realistic tip reproduces both the scaling behavior and the small but significant deviations from scaling observed experimentally.
\end{abstract}

\maketitle
\section{Introduction}
The physical process leading to the emission of electrons from a sharp metallic tip placed at a well defined distance $d$ from a planar electrode depends on $d$. If $d$ is less than about $1$ nm the current is produced by direct tunneling between emitter tip and planar collector.\cite{Simmons} This type of tunnel junction is used, e.g., in Scanning Tunneling Microscopy\cite{Binnig} (STM). At larger distances, the current is dominated by electric-field assisted tunneling of electrons into the region between emitter and collector through the classically forbidden zone.\cite{Simmons,Feenstra,Saenz}
This is the type of junction underlying, e.g., an instrument called the topographiner\cite{topo,AIEP} -- a precursor of STM technology -- and is the elementary building block of recent and less recent developments in nanoelectronics.\cite{Gordon} The quantitative  description of electric-field assisted tunneling, originating within the birth of quantum mechanics, is based on the Fowler-Nordheim equation\cite{O,FN,Simmons,Miller,Mayer} (FN) (see, e.g., Ref. \onlinecite{Forbes} for a summary of recent developments), although recent emitters -- such as, e.g., carbon nanotubes, silicon nanowires, and metallic nanowires-graphene hybrid nanostructures\cite{Spin,Ahmed,Arif} -- as opposed to the ``older ones''\cite{Dyke} -- are almost atomically sharp and thus differ essentially in their geometry from the planar electrode geometry underlying the FN equation. For such field emitters, the potential drop within the tunneling barrier is non-linear and simulations have revealed that the FN equation must be modified to take such non-linearities into account.\cite{Cutler,Fursey}\\ In this paper we systematically measure the current--voltage characteristics of a junction in the electric-field assisted regime, whereby the distance $d$ between tip and collector is varied from a few mm to a few nm.
We observe that the family of curves describing the current $I$ as a function of the two independent variables $V$ \emph{and} $d$ can be ``collapsed'' onto one single scaling curve when $I$ is plotted as a function of the single scaling variable $Vd^{-\lambda}$. This collapsing by a power law of $d$ implies that the physical law governing the flow of current is invariant with respect to changes of the length scale $d$. We argue that the exponent $\lambda$ originates within the solution of the Laplace equation of electrostatics in the vicinity of a singular point and undergoes crossovers to different values depending on whether $d$ is comparable to the total length $L$ of the tip, $d \!\ll\! L$, and, finally, whether $d$ is comparable to the radius of curvature of the tip (see Fig.~\ref{fig1}).
Both the scaling behavior and small but significant deviations from it observed in experiments are reproduced by a model which assumes the electrostatic potential of a realistic tip within the classically forbidden region.
\section{Experimental methods} 
\begin{figure}[t]
\includegraphics{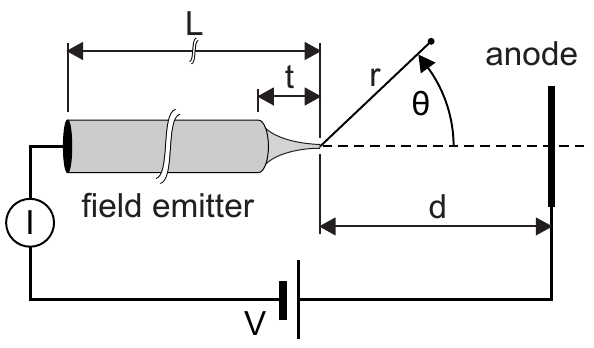}
\caption{\label{fig1} Schematic view of the experimental setup. $L \approx 2$~mm is the total length of the field emitter while $t \approx 250$~$\mu$m is the length of the tip. The current $I$, the voltage $V$, and the distance $d$ are the tunable experimental parameters. ($r,\,\theta$) are the polar coordinates used in the definition of the electrostatic potential $\Phi (d,r,\cos\theta)$.}
\end{figure}

A schematic view of the experimental setup is sketched in Figure~\ref{fig1}. The tip is biased with a negative voltage with respect to the anode, so that field emitted electrons flow from the tip into the anode. Our tips are fabricated starting with a tungsten wire with a few mm length and $250$~$\mu$m diameter. 
The last few hundreds of $\mu$m close to one end of the wire are etched electrochemically to assume a cuspidal profile which, in the final few $\mu$m toward the apex, resembles very much a cone with a full angle of aperture between $\approx \!6^\circ$ and $\approx \!12^\circ$. A rounding of the cone tip limits its sharpness to $\approx \! 5$~nm radius of curvature (at best) but it can be as large as $\approx \! 30$~nm, depending on the details of the tip preparation in an ultra-high vacuum.\cite{AIEP} For the experiments with $d$ in the subnanometer to $\approx \! 2000$~nm range ($d$ being the distance between the apex of the tip and the anode), the collecting plane is a W(110) or a Si(111) single-crystal surface, prepared with standard surface techniques in a  base pressure of $\approx \! 10^{-11}$~mbar. The quality of the tip and the surface topography of the anode are tested by performing standard STM-imaging.\cite{Binnig,Taryl} By mounting the tip onto a piezocrystal, that can move the tip perpendicularly to the surface, the distance $d$ between tip and planar anode can be varied. The approach of the tip to the anode to reach STM-imaging distances (sub-nm) is used to define the origin of the $d$ scale. The STM image of the W(110) surface consists of a few tens of nm wide terraces separated by monoatomic steps, which are known to have a height of $0.2$~nm. This value is used to calibrate $d$ up to $d \!\approx \!2000$~nm, which is the largest displacement we can obtain using our piezocrystal drive. The piezocrystal derived value of $d$ was also  double-checked by an optical sensor device, integrated \emph{ad hoc} into our homemade STM microscope for the purpose of making sure that the piezoreading of $d$ is also reliable at distances much larger that the STM imaging distance. Notice that experiments involving $d$ ranging from several $\mu$m to mm are performed by moving the tip with a mechanically driven positioner. In this case, the origin of the $d$ scale is established by detecting the Ohmic contact between the tip and the anode after the data acquisition.
\vspace{-0.5cm}
\section{Results}
\begin{figure}
\includegraphics{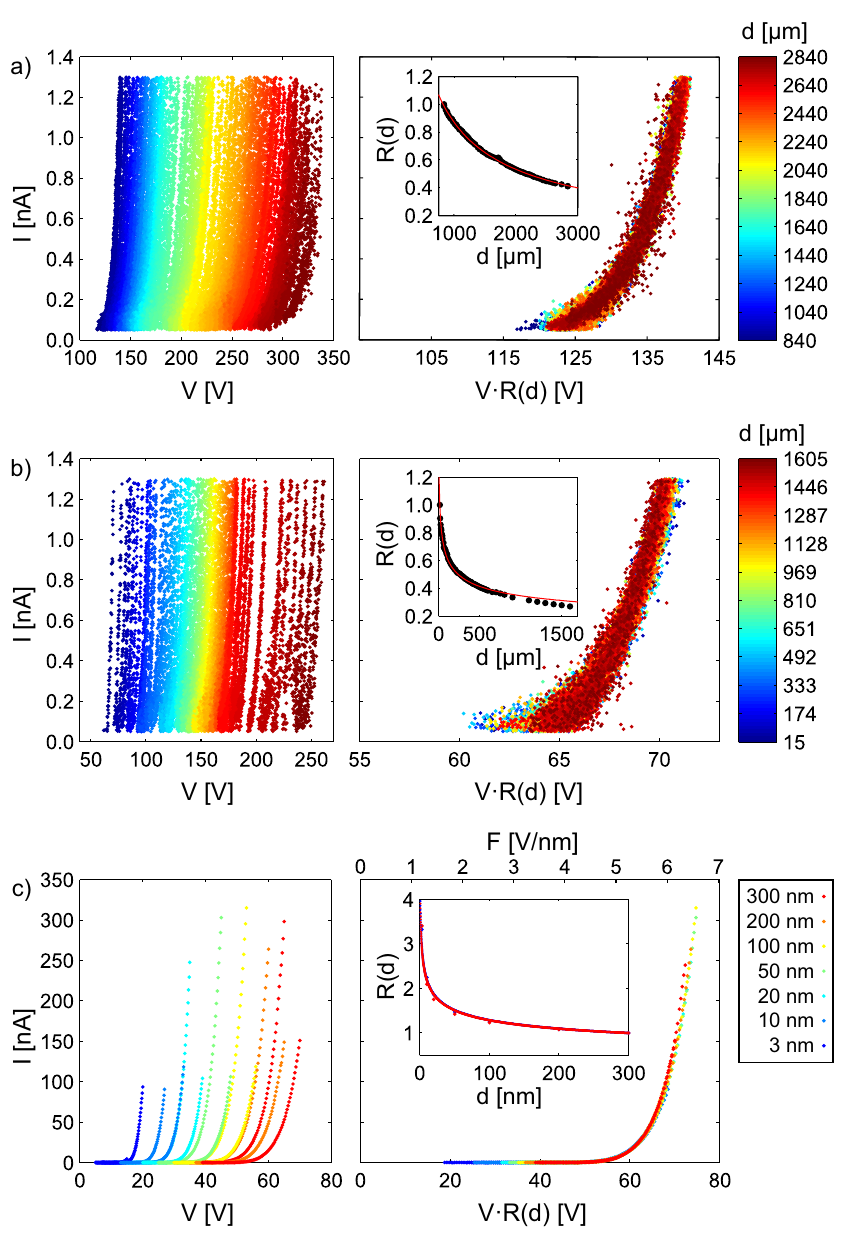}
\caption{\label{fig2}Left: Family of $I$-$V$ curves in the ranges: (a) $0.84$--$2.84$~mm, (b) $15$--$1605$~$\mu$m, and (c) $3$--$300$~nm. Right: The curves on the left are made to collapse onto the curve corresponding to $d \!=\! 0.84$~mm, $14$~$\mu$m, and $300$~nm, respectively, by multiplying the voltage with a number $R(d)$, plotted in the insets on the right. The continuous red curves through $R(d)$ in the insets are a power law $\propto \! 
d^{-\lambda}$ with $\lambda \!=\! 0.71$, $0.27$, and $0.22$, respectively. The variable $d$ is color coded along the vertical bar on the right. The anode used to acquire the data in the top and middle figure was a stainless steel sphere with about a $1$~cm diameter. The alternative horizontal scale in (c) (right) is explained in the text.} 
\includegraphics{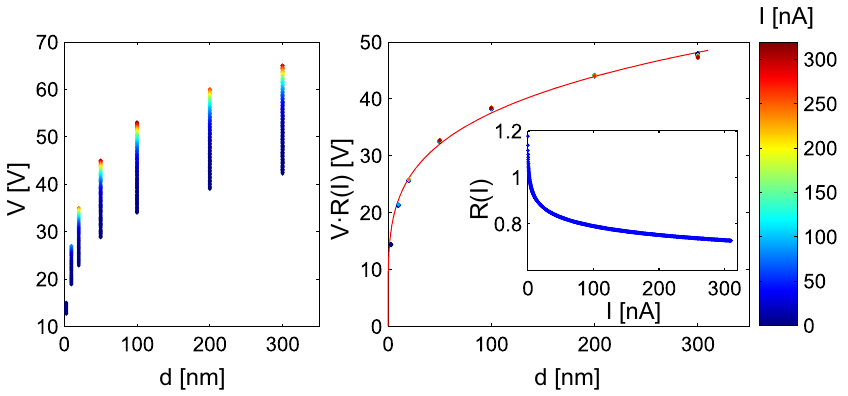}
\caption{\label{fig3}Left: family of $V$-$d$ curves [data taken from Fig.~\ref{fig2}~(c)] in the range $0.2$ to $300$~nA, respectively, each curve corresponding to a given current (color coded along the vertical bar). Right: The family of curves is collapsed onto one single curve corresponding to $I \!=\! 2.3$~nA by multiplying the voltage with a number $R(I)$, plotted in the inset.}
\end{figure}
\subsection{Experimental evidence of scaling} Figure~\ref{fig2} shows scaling plots in three different ranges of values for $d$. On the left, one finds the family of $I$-$V$ characteristics measured within a given $d$ range (color coded along the vertical column). On the right, all $I$ vs $V$ graphs are brought to almost collapse (within experimental noise) onto one single $I$ vs $VR(d)$ reference curve. This means that one entire $I$-$V$ curve at a given distance $d$ is mapped onto a reference $I$-$V$ curve (which can be chosen arbitrarily) by multiplying all voltages by the same number $R(d)$ (scaling factor), plotted in the inset on the right figures. The scaling factor $R(d)$  depends on $d$ but not on $I$ and behaves approximately as a power law $d^{-\lambda}$ [see the red continuous curve through the graphs of $R(d)$]: This means that the current flowing obeys a scaling law of the type $I \!=\! {\cal I}(Vd^{-\lambda})$, i.e., it is a generalized homogeneous function of $V$ and $d$ (Ref. \onlinecite{Stanley}), ${\cal I}(x)$ being the scaling function. The exponent $\lambda$ depends on the range of $d$: it crosses over from $\lambda \!\approx\! 0.71$ [Fig.~\ref{fig2}~(a), $d \!=\! 
850$--$3000$~$\mu$m] to $\approx\! 0.27$ [Fig.~\ref{fig2}~(b), $d \!=\! 15$--$1600$~$\mu$m] to about $0.22$ [Fig.~\ref{fig2}~(c), $d \!=\! 3$--$300$~nm].\\ The same set of experimental data can be represented in a $V$-$d$ diagram as, e.g., in Fig.~\ref{fig3}~(left). In a $V$ vs $d$ plot the distance is varied and the voltage is changed in such a way that the current flowing is kept constant over the whole $V$-$d$ curve. The family of curves in Fig.~\ref{fig3}~(left) can be collapsed onto one single graph [Fig.~\ref{fig3}~(right)] by multiplying the voltage $V$ by a factor $R(I)$, common to all distances, plotted in the inset.
$R(I)$ can be described by a power law $R(I) \!\propto \! I^{-\mu}$ as well, with $\mu \!\approx\! 0.03$--$0.04$, signifying that $d$ depends on one scaling variable, i.e.,  $d \!=\! {\cal D}(V I^{-\mu})$, ${\cal D}(y)$ being the scaling function.
\subsection{The scaling functions} By setting the scaling variable $V d^{-\lambda} \!=$\,const (namely, fixing the current) one obtains $d \!\propto\! V^{1/\lambda}$, i.e., ${\cal D}(y) \!\propto\! y^{1/\lambda}$ or, equivalently, $V \!\propto\! d^{\lambda}$.
One could proceed similarly to obtain ${\cal I}(x)$ from $V I^{-\mu} \!=$\,const. However, being aware that, in practice, it is almost impossible to distinguish a power law ${\cal I}^{-\mu}$ from $\ln(1/{\cal I})$ (Ref. \onlinecite{Id}), when $\mu$ is very small, we prefer to use $V \ln (1/{\cal I})\!=$\,const to obtain the scaling function ${\cal I}(x) \!\propto\! \exp(-1/x)$: This functional dependence is more established in the literature dealing with quantum tunneling \cite{O,FN,Simmons} because it is directly related to the quantum-mechanical transmission coefficient.
\subsection{Experimental determination of the exponent $\lambda$} 
\begin{figure}
\includegraphics{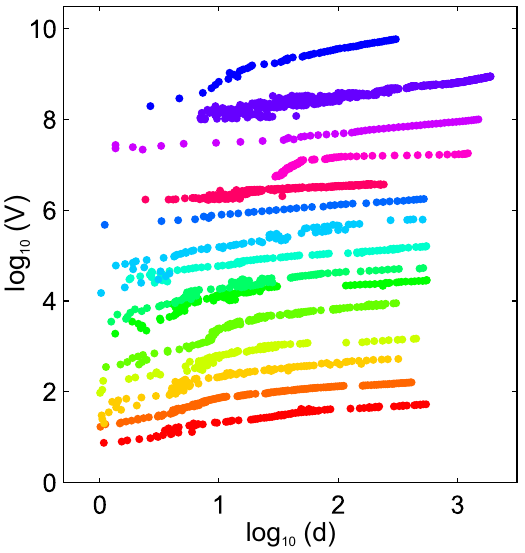}
\caption{\label{fig4}Summary of $V$-$d$ curves at $0.2$~nA in a $\log_{10} V$ vs $\log_{10} d$ plot, $d$ from $\approx \! 3$~nm to $\approx \! 2000$~nm. The curves have been shifted along the vertical scale relative to each other for clarity.} 
\end{figure}

Collapsing of $I$-$V$ curves has been occasionally encountered in experiments [e.g., in the range $0.1$--$0.8$~mm (Ref. \onlinecite{Liu})] as well as a power-law dependence on $d$ with $\lambda \!=\! 0.19$ in the $2$--$14$~$\mu$m range.\cite{Hii} However, the systematic collapsing, accompanied by the emergence of characteristic exponents describing entire sets of graphs over wide ranges of $d$, needs justification.
To avoid the crossover observed when $d$ enters a range comparable to the length $L$ of the emitter, we performed experiments in the range $d \!\ll\! L$ \textbf{(}remarkably, the same crossover appeared in finite-elements simulations of a similar junction\cite{Ed}\textbf{)}.
Figure~\ref{fig4} summarizes experimental $V$-$d$ curves. In a typical experiment, the field-emitter current is set to some prefixed value. The distance $d$ is then varied and the voltage required to keep the current at the prefixed value is measured. In this way, one produces a family of $V$-$d$ curves at selected currents. In the range $d \!\geq\!  10$~nm, the observed linearity (in the $\log$-$\log$ plot) is a manifestation of a  power-law behavior $\propto\! d^{\lambda}$. Most of our data in this range of $d$ give an exponent $\lambda \!\approx\! 0.2 \!\pm\! 0.05$, with some exceptions reaching values up to $\approx\! 0.35$ (see the top graph). This means that the exponent is probably not universal but concentrates around a value of about $0.2$. A systematic downward bending of the graphs in the range $d \!<\! 10$~nm also emerges from Fig.~\ref{fig4}, which indicates a crossover to a possibly different power-law regime with a larger exponent. Thus, these experiments can be summarized as indicating a power-law exponent $\lambda \!\approx\! 0.2\!\pm\!0.05$, in the range $10 \!\lesssim\!d \!\lesssim\! 2000$~nm, with a crossover toward larger values in the range $d \!\lesssim\! 10$~nm.
\subsection{The origin of the exponent $\lambda$} Our working hypothesis for explaining the appearance of $\lambda$ and the measured value is that keeping a constant current is equivalent to keeping a constant electrostatic potential within the classically forbidden region.\cite{Saenz} This is because the electrostatic potential $\Phi(x)$ within the classically forbidden region determines the transmission probability through the classically forbidden region via the quantum-mechanical Gamov exponent, which in its simplest version can be written as $\frac{\sqrt{8m}}{\hbar}\!\int_{x_1}^{x_2}\sqrt{\varphi-\mid \!e\Phi(x)\!\mid}dx$ ($\varphi$ is the work function of the tip, $x$ is the spatial coordinate along the tip axis, and the spatial integration is performed between the classical turning points $x_i$). We search for a hypothetical scaling behavior of the electrostatic potential for the simple case of a tip (set for convenience to zero potential) with conical shape positioned in front of a plane set at a constant potential $V>0$ and at a variable distance $d$. Defining as ($r,\,\theta$) the variables describing the distance from the tip of the cone and, respectively, the polar angle measured with respect to the cone axis (see Fig.~\ref{fig1}), the potential in the vicinity of the tip reads~\cite{Jack} %
\begin{equation}\label{eq1} \Phi (d,r,\cos\theta) \cong V \; A_{\lambda_1}(d) \; r^{\lambda_1} \;
P_{\lambda_1}(\cos\theta) \; .
\end{equation}
Rigorously, Eq.~\eqref{eq1} gives the first term of an infinite sum. 
$P_{\lambda_1} (\cos\theta)$ is a Legendre function of the first kind and of order $\lambda_1$. The exponent $\lambda_1$ is the smallest  positive real number for which the boundary condition $P_{\lambda_1} (\cos\beta) \!=\! 0$ is fulfilled, $\beta$ being the polar angle describing the cone surface ($\pi/2 \!<\! \beta \!<\! \pi$). The denumerable infinite set of non-leading terms are labeled by the numbers $\lambda _i \!>\! \lambda_1$. The coefficients $A_{\lambda_i}(d)$ are determined by the boundary condition $\Phi(d,r\!=\!d/\cos\theta,\cos \theta)\!=\!V$, with $\theta \in [0,\pi/2]$.\\ \emph{Statement.} The coefficients $A_{\lambda_i}$ fulfill the scaling property 
\begin{equation}\label{eq2} 
A_{\lambda_i}(d) = A_{\lambda_i}(1) \; d^{-\lambda_i}\;.
\end{equation}
\emph{Proof.} We notice that $\Phi(1,r/d,\cos\theta)$ fulfills the same Laplace equation and boundary conditions as $\Phi(d,r,\cos\theta)$. As the solution of the Laplace equation fulfilling given boundary conditions is unique, we can set $\Phi(d,r,\cos\theta) \!=\! \Phi(1,r/d,\cos\theta)$. The scaling property Eq.~\eqref{eq2} follows immediately. \hfill $_\blacksquare $ \\
Equation~\eqref{eq2}, combined with Eq.~\eqref{eq1}, states that varying $d$ does not affect the potential $\Phi(r,\theta,d)$ close to the apex of the cone {\it if the applied voltage $V$ is changed in such a way that  $V A_{\lambda_1}$ is independent of $d$}. In virtue of the scaling property Eq.~\eqref{eq2}, $V A_{\lambda _1}\!=$\,const is equivalent to $V \!\propto\! d^{\lambda_1}$, with an exponent defined by the electrostatic boundary condition on the cone surface. 
Indeed, our field emitters end with an almost conical profile,\cite{AIEP} with a full angle of aperture of typically $6^\circ$ to $12^\circ$. This leads to $\lambda_1$ in the range $0.14$--$0.17$,\cite{Jack} i.e., close to the value $\lambda \!\approx\! 0.2$ measured for a large number of field emitters. 
Notice that close to the apex the cone singularity is removed by rounding and/or the formation of a small sphere, with radius of curvature $r_0$ that can vary from about $4$ to $5$~nm at the best to some tens of nm for the bluntest field emitters.\cite{AIEP}
The question arises whether the power-law scaling in Eq.~\eqref{eq2} expected for a cone remains valid for realistic tips and what values of $\lambda_1$ should be used. We point out that the scaling with $d$ of Eqs.~\eqref{eq1}~and~\eqref{eq2} for $r_0 \!\ll\! d \!\ll\! L$ is ``protected'' by Saint-Venant's principle.\cite{Venant} Applied to the present problem, that principle states that if the rounding of the singularity is local enough, the term in Eq.~\eqref{eq1} (containing some cutoff length of the order of $r_0$) remains the leading one.
Only when $d$ approaches values comparable to the radius of curvature can we expect the remaining terms $\lambda_i \!>\! \lambda_1$ to play a role and to produce a crossover to a different solution. We verified the robustness of the conical power-law exponent by explicitly treating the problem of a tip with various typical geometries. For a hyperboloid of revolution, we find that when $d$ is sufficiently large, the exponent $\lambda_1$ is determined by the angle between the axis of the hyperboloid and the asymptotes of the hyperboloid---just as in the conical case. When the plane approaches the tip, the power law with exponent $\lambda_1$ crosses over to a power law with exponent $1$ (typical of a planar capacitor), via a $\log d$ behavior close to the confocal plane.\cite{zuber} We find, extending the work of Miskovsky~\emph{et al.},\cite{miskov} the same $\lambda_1$ and $\lambda_1 \rightarrow 1$ crossover for a plane in the vicinity of a tip with cuspidal shape, which is ultimately the true physical shape of our field emitters up to a few tens of $\mu$m away from the apex. For a plane in the vicinity of a paraboloid of revolution we have $\lambda_1 \!=\! 0$ (meaning a $\log d$ dependence) and no crossover. We have also extended the results of Hall~\cite{hall} and Dyke~\emph{et al.}~\cite{Dyke} to deal with a conical tip terminated by a small sphere and placed in the vicinity of a planar anode at variable $d$: we found the same $\lambda_1$ and
$\lambda_1 \rightarrow 1$ crossover. Finally, we have verified the
$\lambda_1 \rightarrow 1$ crossover numerically and analytically within a simple model mimicking the rounded conical tip with a sequence of small spheres of increasing radius.~\cite{kyri} We point out that our working hypothesis for explaining $\lambda_1$ completely neglects the charge existing within the region of space between the apex of the tip and the planar anode---if this charge is large, one enters the ``screening'' regime, in opposition to the electrostatic regime underlying our hypothesis. The boundary between those regimes is not sharp, but our data seem to indicate that the electrostatic regime dominates in the range of currents used in the present paper. 
In the screening regime, different values for the exponent $\lambda_1$ (Ref. \onlinecite{li}) are expected.
%
\begin{figure}[t]
\includegraphics{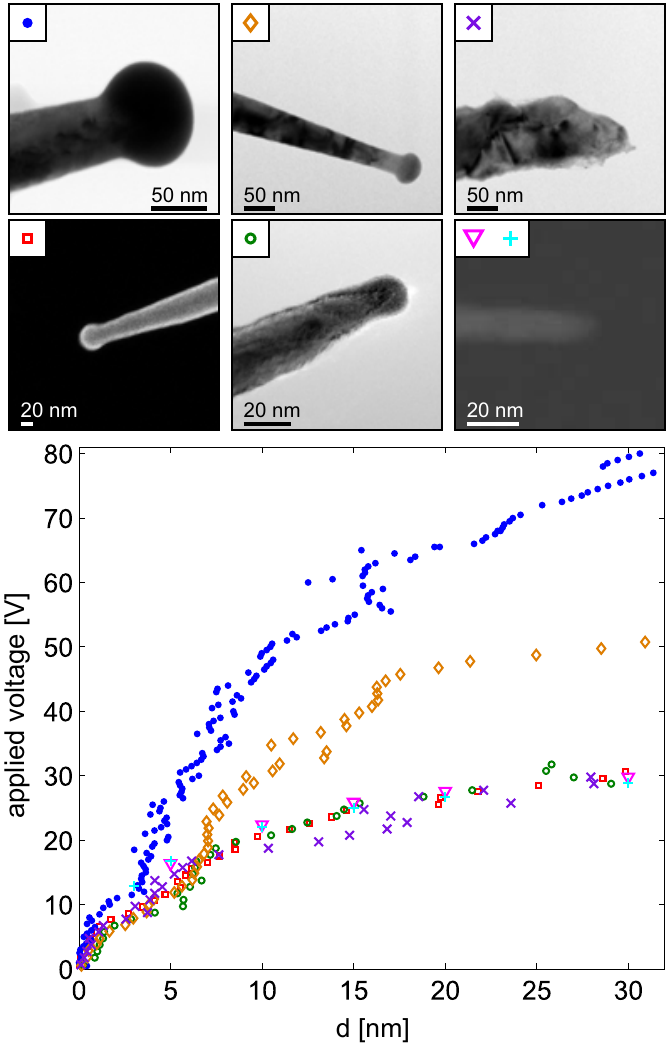}
\caption{\label{fig5}Top shows electron microscopy images of the tips used to obtain the $V$-$d$ curves at $0.2$~nA, the initial section of which is shown at the bottom.} 
\end{figure}
\subsection{The scaling variable $V R(d)$} At any distance, the rounding of the tip introduces a linear component of the potential in the vicinity of the tip apex, which produces a finite electric field at the tip apex $F \! \doteq\! -\partial \Phi/\partial r |_{r \approx r_0}$ pointing along the tip axis. Remarkably, $F$ transforms as $Vd^{-\lambda_1}$ and can, therefore, be associated with the scaling variable $VR(d)$ \textbf{(}see the alternative horizontal scale in Fig.~\ref{fig2}~(c)\textbf{)}. The scaling of the $I$-$V$ characteristics to a reference curve is therefore equivalent to assigning the same field $F$ to junctions with different $d$.
For determining $F$ we have performed $V$-$d$ experiments using tips with various radii of curvature, Fig.~\ref{fig5}, with the aim of detecting the straight, planar-capacitor-like $V$-$d$ section in the vicinity of $d \!=\! 0$.
For microscopy imaging, the tips  were removed from the ultra-high vacuum environment where $V$-$d$ curves were previously taken.
The data of Fig.~\ref{fig5} indicate that, when the radius of curvature is small (typically $5$--$16$~nm, data points $+$,$\times$,$\circ$,$\triangledown$,$\square$), a strong curvature persists, even in the close vicinity of $d \!=\! 0$. For such ``sharp'' field emitters, therefore, we can only conclude that the electric field at the apex is $F \!\geq\! V/d$.
However, our simulations~\cite{kyri} indicate that, with a larger radius of curvature, a larger voltage is required to draw a given current, so that a wider initial section of the graph might emerge, that can be considered straight enough to obtain a reliable value of the electric field from the planar capacitor expression. We fabricated field emitters with larger radii of curvature which allowed us to estimate the electric field $F \!=\! 4.2 \!\pm\! 0.5$~V(nm)$^{-1}$ for a current of $0.2$ nA from a linear fit of the section in the range $d \!\leq\! 10$~nm of the $\bullet$ curve.
\begin{figure}
\includegraphics{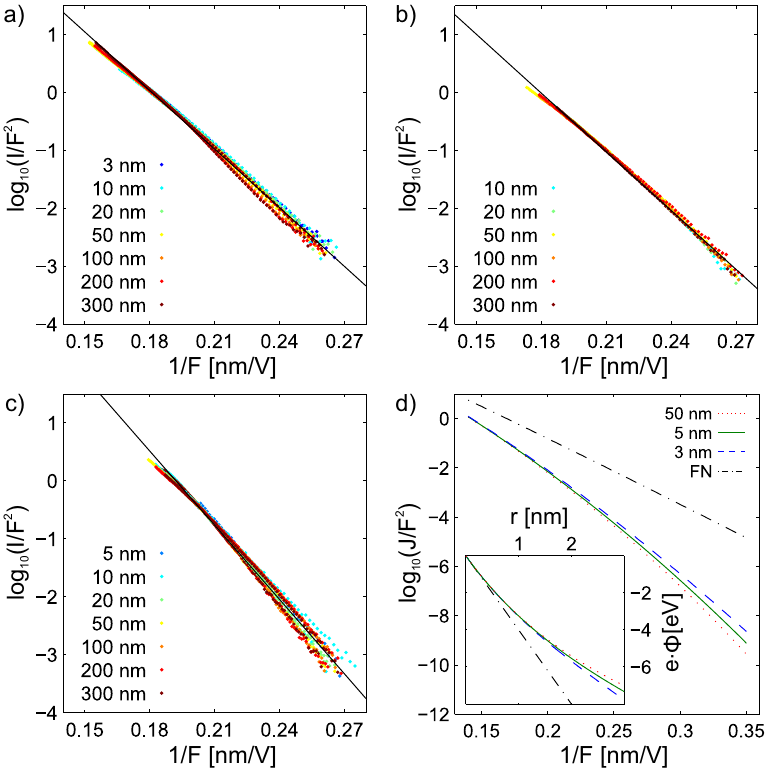}
\caption{\label{fig6}(a)--(c) Experimental $\log_{10}(I/F^2)$ vs $1/F$ diagram for three different tips in the $3$--$300$~nm range. 
$I$ is given in nA and $F$ in V\,(nm)$^{-1}$. The straight lines are linear fits to the experimental data. The data of Fig.~\ref{fig4}~(a) are those of Fig.~\ref{fig2}~(c). (d) Computed $\log_{10}(J/F^2)$ vs $1/F$ for a field emitter with a radius of curvature of $4$~nm at different $d$.
The field-emission current density $J$ was computed as described in Ref.~\onlinecite{kyri}.
The dashed-dotted line is the standard Fowler-Nordheim plot\protect.~\cite{Miller} In the inset, the potential energy $e \Phi$ is plotted as a function of the distance $r$ from the the tip, showing an ``upward'' curvature that effectively increases the width of the tunneling barrier with increasing $d$ [electric field at the apex is set to $4$~V\,(nm)$^{-1}$].}
\end{figure} 
\subsection{The tunneling barrier} Figures~\ref{fig6}~(a) to (c) plot experimental $I$-$V$ curves, taken in the $3$--$300$~nm range, within a $\log_{10}(I/F^2)$ vs $1/F$ diagram.  These plots reveal that
\begin{inparaenum}[\upshape(\itshape\!i\upshape)]
	\item for a given field the current acquires approximately the same value, independently of $d$, reiterating the scaling behavior reported in Fig.~\ref{fig2}. The data also reveal
	\item a small but detectable downward curvature of the graphs accompanied by
	\item a small but detectable non-collapsing in the range of small currents.
\end{inparaenum}
Figure~\ref{fig6}~(d) reports simulations of the tunneling process at variable $d$ (Ref. \onlinecite{kyri}). The tip was simulated with a set of small spheres aligned along the direction perpendicular to the planar anode and with increasing radius of curvature. The topmost sphere determines the radius of curvature, the following ones with increasing diameter mimic the angle of aperture of the cone. The plot of Fig.~\ref{fig6}~(d) shows the computed current {\it density} $J$ versus electric field in a $\log_{10}(J/F^2)$ vs $1/F$ diagram, for different distances $d$. Evident from Fig.~\ref{fig6}~(d) are the three aspects observed experimentally in Figs.~\ref{fig6}~(a) to (c), namely, that
\begin{inparaenum}[\upshape(\itshape\!i\upshape)]
	\item the graphs for different $d$ almost coincide (scaling),
	\item the graphs are slightly curved downward, in agreement with the results of Refs.~\onlinecite{Cutler, Fursey} for tips with similar radii of curvature, and that
	\item the downward curvature of the graphs toward small electric fields increases with distance $d$.
\end{inparaenum}
The origin of these three aspects seems to be the behavior of the electrostatic potential in the vicinity of the tip; see the inset of Fig.~\ref{fig6}~(d). In fact, the inset shows that for a given electric field $F$
\begin{enumerate}[\upshape(\itshape\!i\upshape)]
	\item  the spatial dependence of the electrostatic potential is almost independent of $d$,
	\item the spatial dependence of the potential has an upward curvature that effectively widens the classically forbidden region, as pointed out in Refs.~\onlinecite{Cutler,Fursey} and that
	\item for increasing $r$ the $d$-dependence of $\Phi(r)$ becomes more pronounced.
\end{enumerate}
A quantitative comparison between the experimental data of Figs.~\ref{fig6}~(a) to (c) and the simulation results of Fig.~\ref{fig6}~(d) can be done on the base of the average slope of the graphs within the range of fields $0.16 \!<\! 1/F \!<\! 0.3$ ($1/F$ in units of nm\,V$^{-1}$). We speak of ``average'' slope because the graphs in Figs.~\ref{fig6}~(a) to (c) as welle as the simulations in Refs.~\onlinecite{Cutler,Fursey} and Fig.~\ref{fig6}~(d) have a slight downward curvature that makes the slope dependent on $F$. We use, as unit for the slope, the original expression by FN for an exact triangular barrier:~\cite{FN,Miller} $-\frac{4 \log_{10}({\rm e})}{3} \frac{\sqrt{2m}}{e\hbar} \varphi^{3/2}$ \textbf{(}$\varphi=4.5$~eV in the present paper~\cite{Feenstra}\textbf{)}. The average (multiplicative) slope correction factor derived from Figs.~\ref{fig6}~(a) to (c) falls in the range $1.2$--$1.5$, to be compared with the range $1.15$--$1.44$ for $r_0 \!=\! 8$--$4$~nm (Ref. \onlinecite{Fursey}), $1.43 \!\pm\! 0.1$ for $r_0 \!=\! 4$~nm [our simulations, Fig.~\ref{fig6}~(d)], and $\approx\! 0.95$, recommended for the triangular FN tunneling junction with image potential correction.~\cite{Miller,Forbes}
\section{Summary} We have provided experimental evidence for the scale invariance of a tunnel junction with respect to changes of a characteristic length from nm to mm. We have also provided an explanation of this phenomenon in terms of electrostatics of sharp boundaries. A simple model of electric-field assisted tunneling using a ``realistic'' tip geometry reproduces the essential features observed experimentally. The accuracy of the data collapsing is remarkably high, taking into account that the notion of scale invariance is certainly better known in sciences~\cite{Stanley} other than solid-state devices. The (almost) scale invariant electric-field assisted tunneling regime described in the present paper is essentially different from the direct-tunneling (STM) regime,~\cite{Binnig} achieved when $d$ is in the sub-nm range, marked by the appearance of a characteristic length~\cite{Simmons} and by almost linearity of $I\!-\!V$ characteristics.~\cite{Feenstra}
\section*{ACKNOWLEDGMENTS}
 We thank Thomas B\"ahler for technical assistance, the microscopy center of ETHZ (EMEZ) for some of the tip images, the Swiss National Science Foundation and ETH Zurich for financial support, R.~Forbes for suggesting to D.~P. the notion of ``slope correction factor'', and G.M.~Graf for the proof of the theorem. Ar.~A. acknowledges financial support by the Welch Foundation (Grant No. A-1678).

\end{document}